\documentclass[article,nojss,nofooter]{jss}

\usepackage{amsmath}
\usepackage{booktabs}
\usepackage{graphicx}
\usepackage{url}

\newcommand{\class}[1]{\code{#1}}

\author{Ihor Kendiukhov\\University of T\"ubingen}
\Plainauthor{Ihor Kendiukhov}

\title{Ergodicity Library: A \proglang{Python} Toolkit for Stochastic-Process Simulation, Time-Average Diagnostics, and Agent-Based Experiments}
\Plaintitle{Ergodicity Library: A Python Toolkit for Stochastic-Process Simulation, Time-Average Diagnostics, and Agent-Based Experiments}
\Shorttitle{Ergodicity Library in \proglang{Python}}

\Abstract{
\pkg{ergodicity} is an open-source \proglang{Python} library for computational work on stochastic dynamics, with particular emphasis on non-ergodicity, time-average behavior, heavy-tailed processes, and decision making under uncertainty. The package brings together three layers that are often split across ad hoc scripts: process definitions and simulators, analysis and fitting tools, and agent-based experimentation. This article documents the implemented software rather than presenting new stochastic theory. We describe the package architecture, the supported process families, the analysis workflow, and the practical boundaries of the current implementation. We also provide fully reproducible examples covering heavy-tailed ensemble spread, multiplicative Levy growth diagnostics, adaptive-memory mean reversion, preasymptotic fluctuation analysis, and partial stochastic differential equation simulation. The package is positioned as an integration layer on top of the scientific \proglang{Python} stack, reducing the amount of glue code required to move from process specification to diagnostics and comparative experiments.
}

\Keywords{research software, stochastic processes, heavy tails, ergodicity, time averages, \proglang{Python}}
\Plainkeywords{research software, stochastic processes, heavy tails, ergodicity, time averages, Python}

\Address{
  Ihor Kendiukhov\\
  University of T\"ubingen\\
  T\"ubingen, Germany\\
  E-mail: \email{kenduhov.ig@gmail.com}
}

\begin{document}

\section[Introduction: Computational stochastic workflows in Python]{Introduction: Computational stochastic workflows in \proglang{Python}}\label{sec:intro}

Research on stochastic systems often requires chaining together tasks that are usually scattered across notebooks and one-off scripts: simulating candidate processes, estimating summary statistics, comparing time and ensemble behavior, fitting distributional families, and iterating on decision rules. In practice, the fragmentation cost is substantial. General-purpose scientific libraries provide strong numerical primitives, but they do not by themselves impose a coherent workflow for ergodicity-sensitive questions \citep{harris2020numpy, virtanen2020scipy, meurer2017sympy, hunter2007matplotlib, millman2011}. Reproducibility guidance in computational science has repeatedly highlighted the same failure mode: the logic of an analysis becomes distributed across disconnected scripts, environment assumptions, and undocumented plotting steps \citep{peng2011, wilson2014, wilson2017, benureau2018, katz2016software, lamprecht2020fair}.

\pkg{ergodicity} was developed to reduce that integration burden \citep{kendiukhov2025repo, kendiukhov2024book}. The library is motivated by questions where time averages and ensemble expectations can diverge, especially under multiplicative or heavy-tailed dynamics \citep{peters2016evaluating, peters2019ergodicity}. Rather than targeting a single model class, it organizes simulation, diagnostics, and decision-oriented experimentation around a shared process abstraction.

This manuscript is a software paper in the \emph{Journal of Statistical Software} sense. The goal is to document what the package implements, how it is organized, and what kinds of computational studies it supports. The emphasis is on software design, reproducible examples, and the relationship between the package and the surrounding scientific \proglang{Python} ecosystem.

\section{Package scope and design goals}\label{sec:scope}

The current implementation is best understood as an integrated stochastic-computing toolkit with three explicit design goals.

\begin{enumerate}
\item Provide reusable abstractions for a broad family of stochastic processes rather than isolated simulators.
\item Couple simulation closely to diagnostics such as fitting, preasymptotic analysis, and symbolic or numerical helper routines.
\item Support experiments in which stochastic dynamics and adaptive decision rules interact in the same software environment.
\end{enumerate}

These goals lead to a layered structure with three main package areas: process simulation, analytical tools, and agents. Table~\ref{tab:architecture} summarizes the high-level architecture.

\begin{table}[t!]
\centering
\begin{tabular}{p{0.22\linewidth}p{0.31\linewidth}p{0.39\linewidth}}
\toprule
Component & Main contents & Role in a research workflow \\
\midrule
\code{ergodicity.process} & Process classes, stochastic increments, multiplicative and memory-dependent variants, constructor helpers & Define stochastic hypotheses and generate trajectories under alternative assumptions \\
\code{ergodicity.tools} & Fitting, evaluation, preasymptotics, symbolic helpers, multiprocessing, automation, partial stochastic differential equation simulation & Quantify behavior, compare model families, and automate repeated analyses \\
\code{ergodicity.agents} & Utility-based agents, portfolio abstractions, agent pools, evolutionary optimization, machine-learning-oriented routines & Explore decision rules and interacting adaptive systems built on stochastic environments \\
\bottomrule
\end{tabular}
\caption{High-level package architecture.}
\label{tab:architecture}
\end{table}

A central design choice is the use of a shared process hierarchy. User-facing simulation workflows are exposed through related classes such as \class{Process}, \class{ItoProcess}, \class{NonItoProcess}, and custom-process scaffolding. This allows common operations such as simulation, plotting, and summary extraction to be reused across multiple stochastic families while still supporting model-specific increment definitions.

\section[Processes, diagnostics, and software interfaces]{Processes, diagnostics, and software interfaces}\label{sec:software}

\subsection{Process families}

The process layer covers a wide span of model families. In the current codebase, these include Brownian and Wiener-type models, Levy-stable processes, Poisson-driven dynamics, multiplicative geometric variants, memory-dependent processes, fractional or multifractional variants, and selected multivariate process classes. The purpose is not to claim exhaustive coverage of stochastic modeling, but to provide a practical working set that spans additive and multiplicative dynamics, light-tailed and heavy-tailed increments, and memoryless versus adaptive behavior \citep{oksendal2003, morters2010, higham2001, chambers1976, mandelbrot1968, gillespie1977}.

\subsection{Analytical and diagnostic tooling}

The tools layer is where the package differentiates itself from a minimal simulator. In addition to standard numerical summaries, the package includes distribution-fitting utilities, evaluation helpers, symbolic support routines, preasymptotic diagnostics, automation and multiprocessing wrappers, and a partial stochastic differential equation simulator. This arrangement is valuable when a user needs to move quickly from a new process specification to a diagnostic comparison without re-implementing the surrounding infrastructure.

\subsection{Agents and decision-oriented routines}

The agents layer extends the library beyond passive trajectory generation. It contains utility-based agents, portfolio abstractions over process ensembles, interacting agent pools, and evolutionary or machine-learning-oriented optimization routines. These capabilities connect the stochastic-process layer to decision-making workflows and to broader agent-based modeling practice \citep{bonabeau2002, grimm2005}. In the current package, however, this is also the part most sensitive to optional dependency complexity.

\subsection{A minimal interface example}

A typical workflow starts from process construction and ends with an ensemble diagnostic. The following abbreviated example is representative of the package API.

\begin{Code}
from ergodicity.process.basic import BrownianMotion
from ergodicity.process.multiplicative import GeometricLevyProcess

bm = BrownianMotion(drift=0.0, scale=1.0)
glp = GeometricLevyProcess(alpha=1.55, beta=0.2, scale=0.35, loc=0.02)

brownian = bm.simulate(t=3.0, timestep=0.01, num_instances=240,
    plot=False, save=False)
heavy_tailed = glp.simulate(t=4.0, timestep=0.01, num_instances=360,
    plot=False, save=False)
\end{Code}

At this level the package aims to keep the workflow consistent across process classes. Downstream diagnostics can then consume the same simulation outputs without reshaping the entire analysis around each model family.

\section{Reproducible examples}\label{sec:examples}

All figures in this section are generated by the standalone replication script \code{article.py} included with the submission. The script uses the packaged source snapshot and writes all graphics into the local \code{figures/} directory. The examples were selected to show software functionality rather than to serve as standalone scientific claims.

\subsection{Gaussian versus heavy-tailed ensemble spread}

\begin{figure}[t!]
\centering
\includegraphics[width=0.98\linewidth]{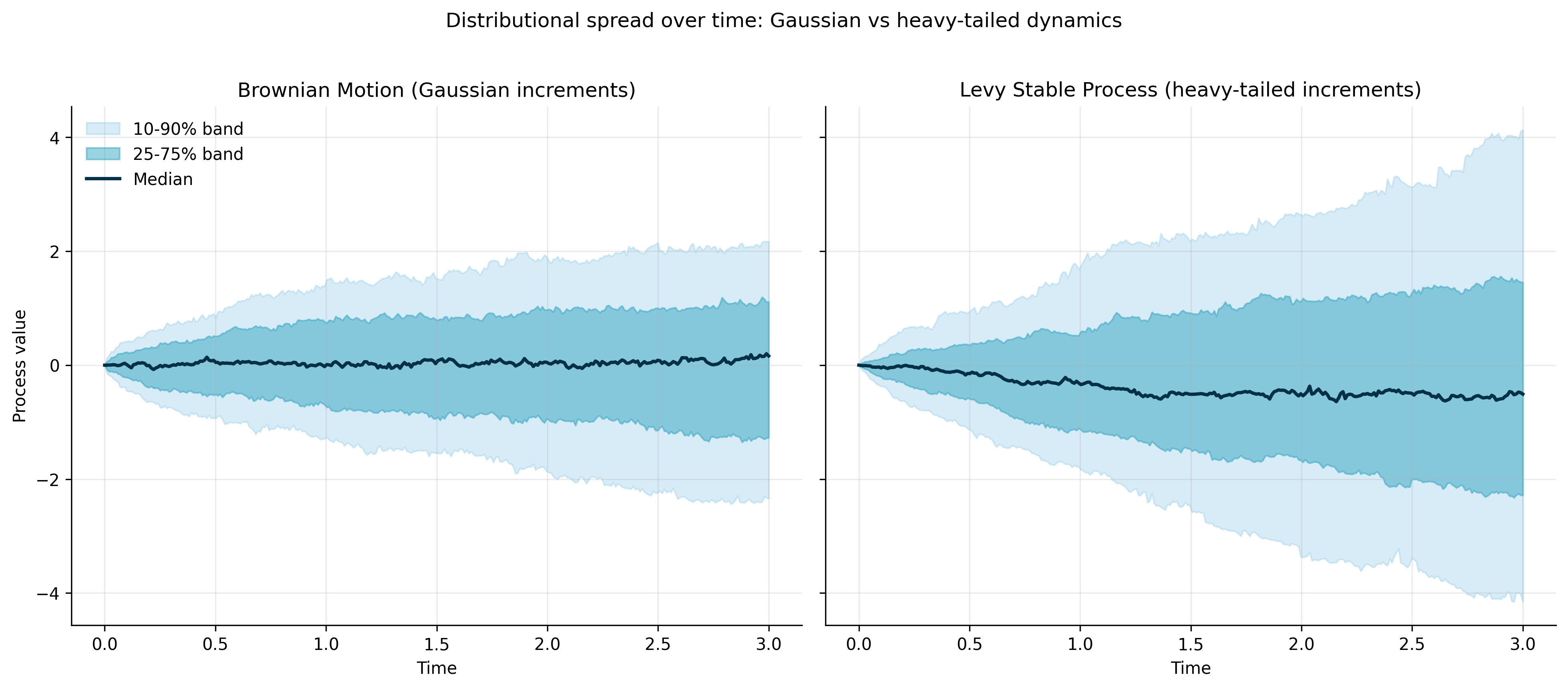}
\caption{Quantile fan charts for Brownian and Levy-stable ensembles. The heavy-tailed process produces a broader and less regular spread geometry over the same horizon.}
\label{fig:heavytail}
\end{figure}

Figure~\ref{fig:heavytail} compares ensemble fan charts for Brownian motion and a Levy-stable process under comparable discretization. The point is not simply that the heavy-tailed model is more variable. More importantly, the geometry of the quantile bands becomes less regular and less well represented by Gaussian intuition, which matters for downstream diagnostics and model selection \citep{mandelbrot1963, bouchaud1990, cont2001}.

\subsection{Heavy-tailed multiplicative growth diagnostics}

\begin{figure}[t!]
\centering
\includegraphics[width=0.98\linewidth]{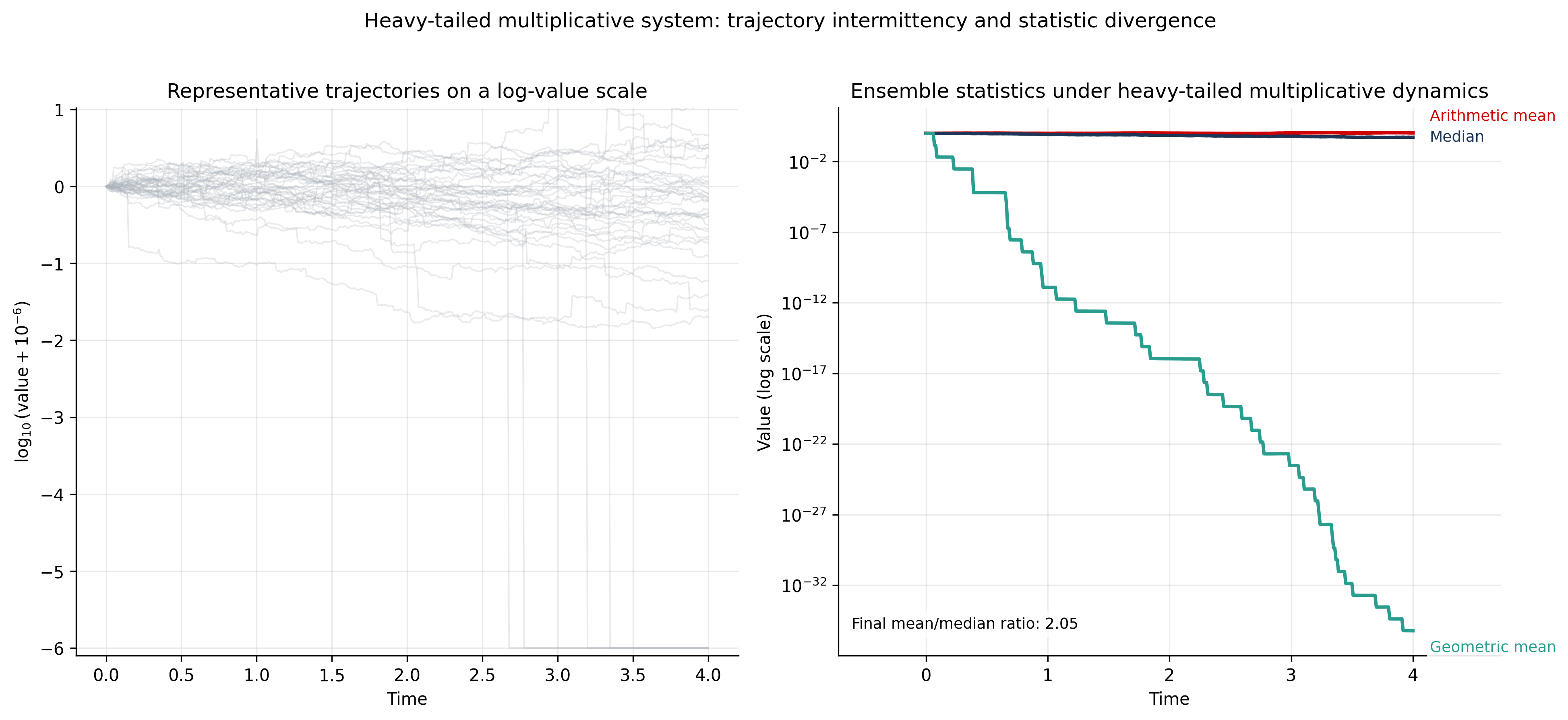}
\caption{Trajectory intermittency and summary-statistic divergence for a Geometric Levy process.}
\label{fig:geolevy}
\end{figure}

The multiplicative example in Figure~\ref{fig:geolevy} is built from a Geometric Levy process. The left panel shows trajectory intermittency on a log scale; the right panel compares arithmetic mean, median, and geometric mean. For ergodicity-sensitive analyses, this gap between ensemble summaries is part of the software story: the package makes it easy to compute and visualize the finite-sample separation between statistics that coincide only under more restrictive assumptions \citep{peters2016evaluating, peters2019ergodicity}.

\subsection{Adaptive memory in mean-reverting dynamics}

\begin{figure}[t!]
\centering
\includegraphics[width=0.92\linewidth]{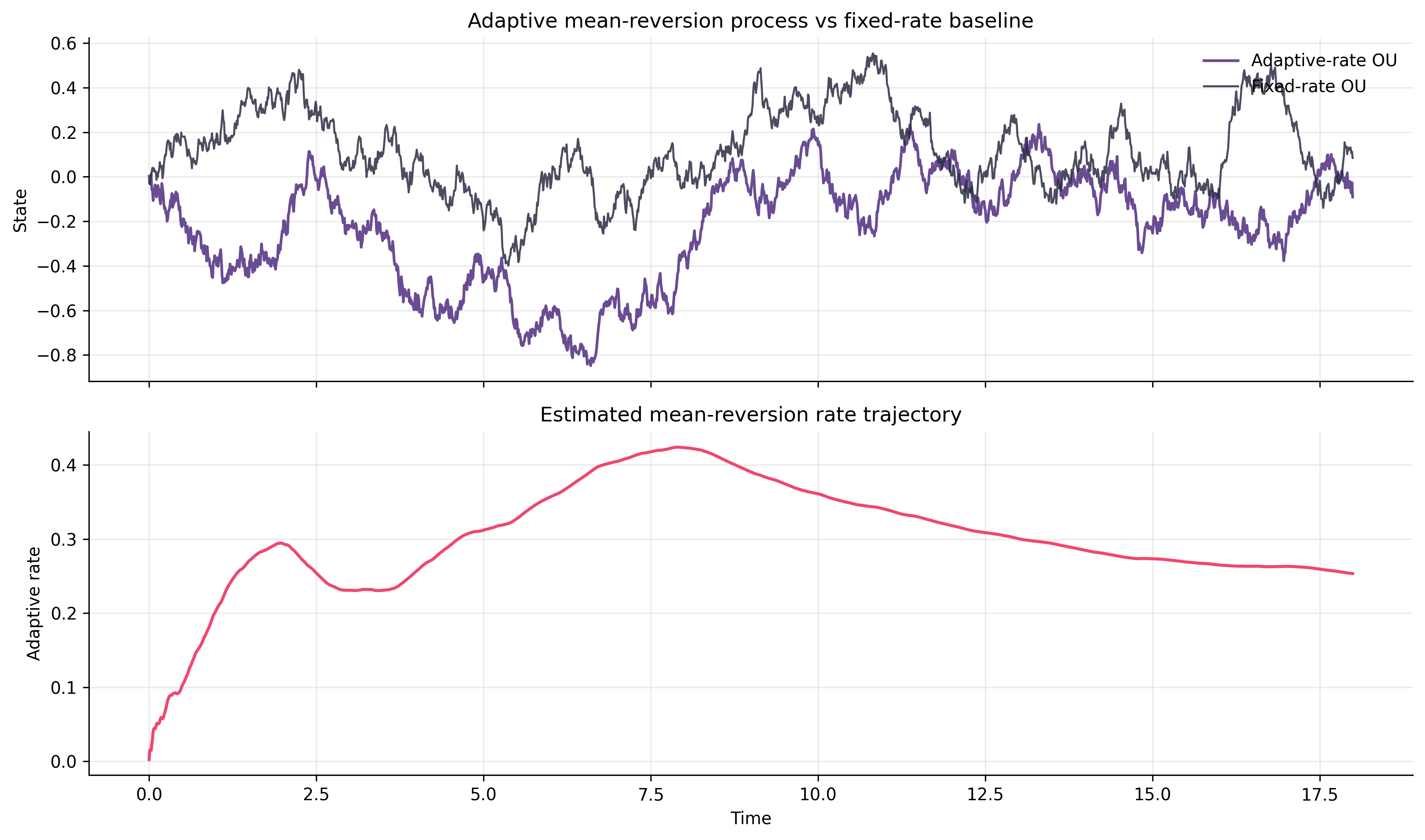}
\caption{Adaptive-rate Ornstein--Uhlenbeck dynamics compared with a fixed-rate baseline.}
\label{fig:adaptiveou}
\end{figure}

Figure~\ref{fig:adaptiveou} illustrates a memory-dependent Ornstein--Uhlenbeck variant against a fixed-rate baseline. The top panel shows the state trajectories, while the lower panel records the evolving mean-reversion rate. This example demonstrates how the package can express models in which an effective parameter adapts to path history instead of remaining fixed.

\subsection{Preasymptotic diagnostics}

\begin{figure}[t!]
\centering
\includegraphics[width=0.98\linewidth]{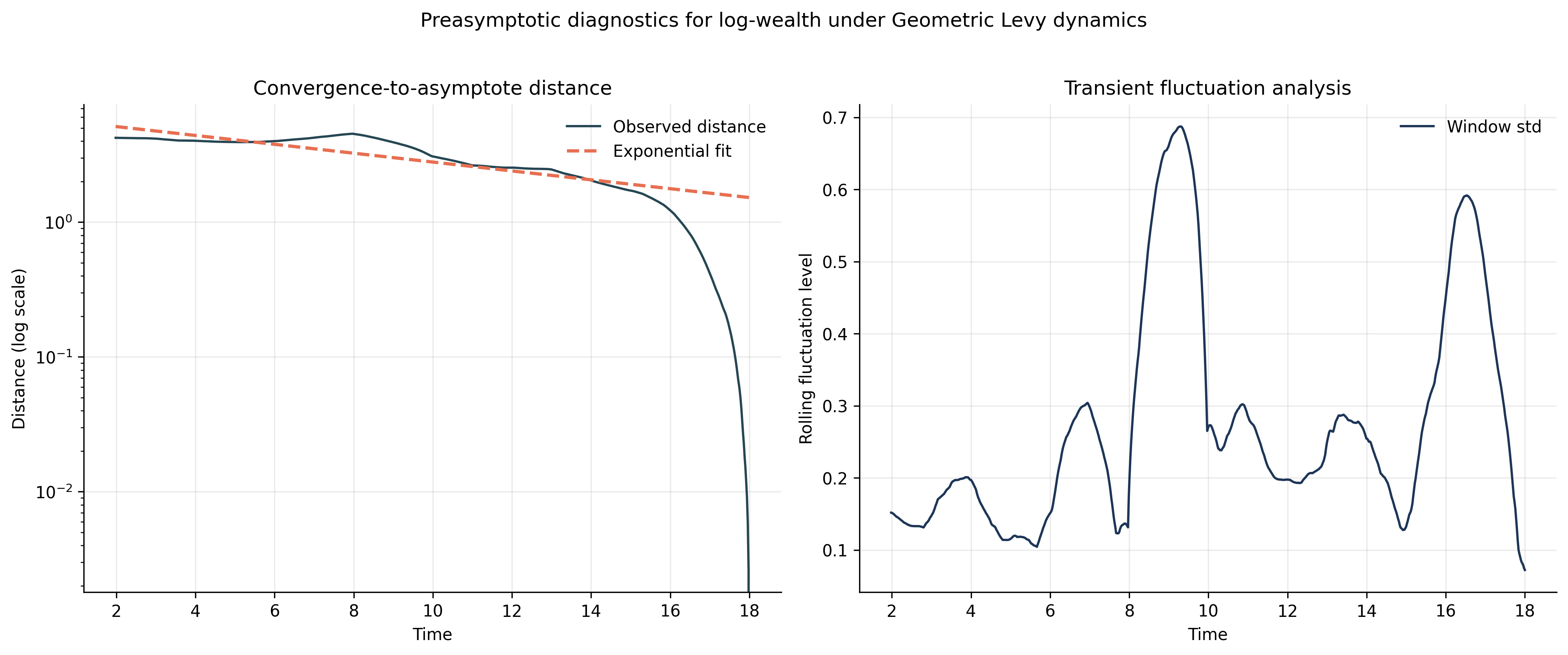}
\caption{Preasymptotic convergence and fluctuation diagnostics for log-wealth under Geometric Levy dynamics.}
\label{fig:preasym}
\end{figure}

Preasymptotic behavior is central when finite-time observations differ materially from asymptotic expectations. Figure~\ref{fig:preasym} shows two diagnostics computed from log-wealth under Geometric Levy dynamics: distance to an estimated asymptote and a rolling fluctuation measure. The package exposes this kind of analysis through the \code{ergodicity.tools.preasymptotics} module, allowing exploratory checks before stronger long-run conclusions are drawn.

\subsection{Partial stochastic differential equation simulation}

\begin{figure}[t!]
\centering
\includegraphics[width=0.98\linewidth]{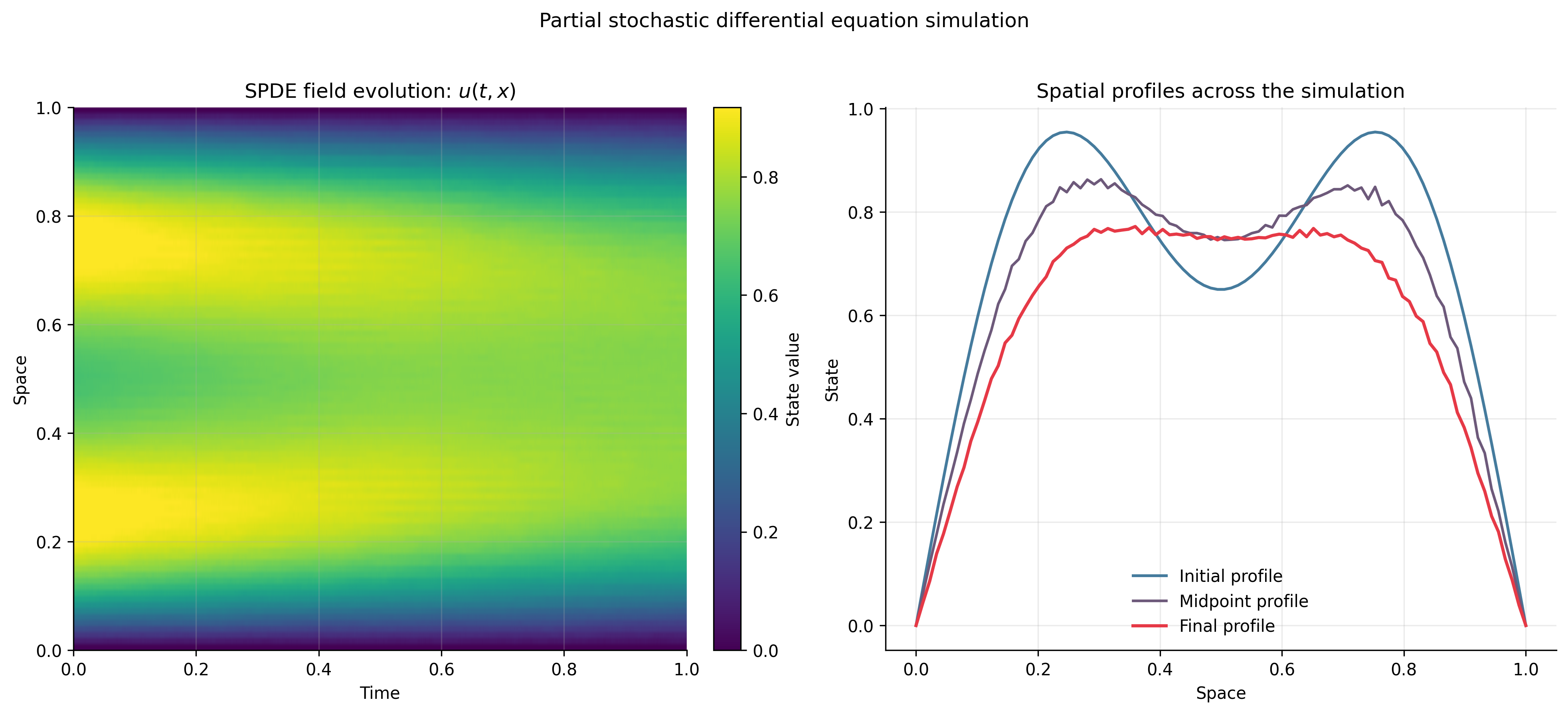}
\caption{Partial stochastic differential equation simulation with a contour view and spatial boundary profiles.}
\label{fig:spde}
\end{figure}

The final example broadens the scope beyond one-dimensional trajectories. Figure~\ref{fig:spde} shows an SPDE simulation with a space--time contour view and initial-versus-final spatial profiles. This demonstrates that the library can also support field-style stochastic computations rather than only scalar time series.

\section[Position in the Python ecosystem]{Position in the \proglang{Python} ecosystem}\label{sec:ecosystem}

\pkg{ergodicity} is not a replacement for the scientific \proglang{Python} stack. Instead, it is an application-layer package built on top of that stack. Numerical arrays and vectorized operations rely on \pkg{NumPy}; optimization, probability distributions, and related routines build on \pkg{SciPy}; symbolic manipulations rely on \pkg{SymPy}; and figures rely on \pkg{Matplotlib} \citep{harris2020numpy, virtanen2020scipy, meurer2017sympy, hunter2007matplotlib}. The value added by \pkg{ergodicity} is not a lower-level numerical primitive. It is the integration of process definitions, diagnostics, and decision-oriented experiments into one package structure.

This distinction matters for how the package should be evaluated. The most direct alternative to \pkg{ergodicity} is not a single competitor package but a collection of project-specific scripts glued together from complementary libraries. Compared with such ad hoc workflows, \pkg{ergodicity} offers three concrete advantages:

\begin{enumerate}
\item a shared stochastic-process abstraction across multiple model families,
\item analysis tools that can consume the corresponding outputs without extensive reshaping, and
\item optional agent-based extensions that use the same process environment.
\end{enumerate}

The package also has clear boundaries. It does not claim to be the fastest available simulator for every supported process, nor does it claim that all optional components are equally mature. The main software contribution is integration breadth, not benchmark dominance.

\section{Availability, testing, and replication}\label{sec:availability}

The submission bundle contains four key elements: the JSS manuscript source, the JSS manuscript PDF, a current source snapshot of the package, and a standalone replication script. The packaged source snapshot is a cleaned copy of the local project tree used for this submission. The replication script imports the snapshot directly so that figure generation does not depend on the state of an external checkout.

The package metadata currently declare version \code{0.3.19}, \proglang{Python}~\code{>= 3.10}, and an MIT license. The primary public repository is the GitHub project \code{ergodicity\_library}\footnote{\url{https://github.com/Kendiukhov/ergodicity_library/}}. Following software-citation and FAIR-oriented guidance, the manuscript cites both the public repository and the surrounding software-literature context \citep{katz2016software, lamprecht2020fair}.

At submission time, the packaged regression test suite is still small, but it is no longer empty. A representative validation command for the current package state is:

\begin{Code}
PYTEST_DISABLE_PLUGIN_AUTOLOAD=1 \
PYTHONPATH=software_source/ergodicity_library \
pytest -q software_source/ergodicity_library/tests/test_multivariate_langevin.py
\end{Code}

In the audited submission environment this command passes. The figure replication command is similarly direct:

\begin{Code}
python article.py
\end{Code}

This writes the manuscript figures into \code{figures/}. The current package still has two practical caveats. First, optional machine-learning paths inside \code{ergodicity.agents} depend on heavier third-party stacks and are more environment-sensitive than the core process and tools layers. Second, numerical conclusions remain sensitive to discretization, horizon choice, and ensemble size, especially for heavy-tailed multiplicative models.

\section{Conclusion}\label{sec:conclusion}

\pkg{ergodicity} provides a broad software environment for stochastic-process simulation, time-average diagnostics, and agent-based experimentation in \proglang{Python}. Its main contribution is not a single new algorithm. Instead, it is the integration of process abstractions, diagnostic tools, and decision-oriented workflows in a package that can be inspected, extended, and reproduced.

For JSS readers, the relevant result is software-level: users can move from process specification to comparative diagnostics and worked computational examples within one consistent package structure. The examples in this article demonstrate that this approach is already useful for heavy-tailed spread analysis, multiplicative growth diagnostics, adaptive-memory processes, preasymptotic analysis, and partial stochastic differential equations. Future work should focus on expanding the test suite, simplifying optional dependency boundaries, and adding more domain-specific benchmark examples without changing the core architecture.

\bibliography{refs}

\end{document}